# Dynamic Power Reduction in a Novel CMOS 5T-SRAM for Low-Power SoC

Hooman Jarollahi, EIT, Student Member, IEEE, and Richard F. Hobson, P. Eng.

School of Engineering Science, Simon Fraser University, Burnaby, B.C., Canada, V5A 1S6

hjarolla@sfu.ca

**Abstract** — *This paper addresses a novel five-transistor (5T) CMOS SRAM design with high performance and reliability in 65nm CMOS, and illustrates how it reduces the dynamic power consumption in comparison with the conventional and low-power 6T SRAM counterparts. This design can be used as cache memory in processors and low-power portable devices. The proposed SRAM cell features ~13% area reduction compared to a conventional 6T cell, and features a unique bit-line and negative supply voltage biasing methodology and ground control architecture to enhance performance, and suppress standby leakage power.*

**Keywords:** Five-transistor SRAM; low-power low-area SRAM; cache memory; standby and dynamic power reduction.

## 1 INTRODUCTION

Today's microprocessor chips consist of cache memories and computing cores. It is predicted that cache memories may reach 90% of the chip area in some applications by 2013 [4]. In addition, cache memories consume a significant portion of the power budget in SoC applications [3]. This is particularly important in portable and battery-powered electronics such as cellular phones, PDAs, wireless, and low-power biomedical devices since dynamic and standby leakage power determine the battery life. With recent aggressive growth of technology scaling, standby leakage power is increased nearly five times each technology generation while active power remains constant [3]. Also, process variations and hence performance fluctuations are widely noticed in 65nm and beyond in CMOS technologies [5]. *Five-transistor Static Random Access Memories* (5T SRAMs) are attractive due to their advantage in area and power efficiency compared to 6T SRAMs [1][2][8][9][16]. Research in the past on this type of memory has been mostly focused on improving performance and stability while maintaining the promised area saving in a particular technology node. On the other hand, with continuous scaling down of CMOS transistors, new techniques have been developed in 6T SRAMs such as Dynamic Standby Mode [4][12], DRV method [3], and well biasing, some of which are summarized in [3] and [4]. Therefore, in order to suppress leakage power consumption and combat performance fluctuations due to process variations, the previous research in 5T SRAMs such as [8] and [9], can no longer compete with current 6T SRAMs and that is why 6T SRAMs are still predominantly used in current systems.

In [1], standby power reduction has been described for the new 5T SRAM design. In this paper, an improved low-power design with the focus on its dynamic power reduction advantage is addressed. The new 5T SRAM cell with dual grounds (5TSDG) features a novel bit-line biasing technique, and guarantees operation under all process variations and temperatures while taking benefit of area reduction. In addition, 5TSDG has an improved performance compared to previous research in [2][8][9].

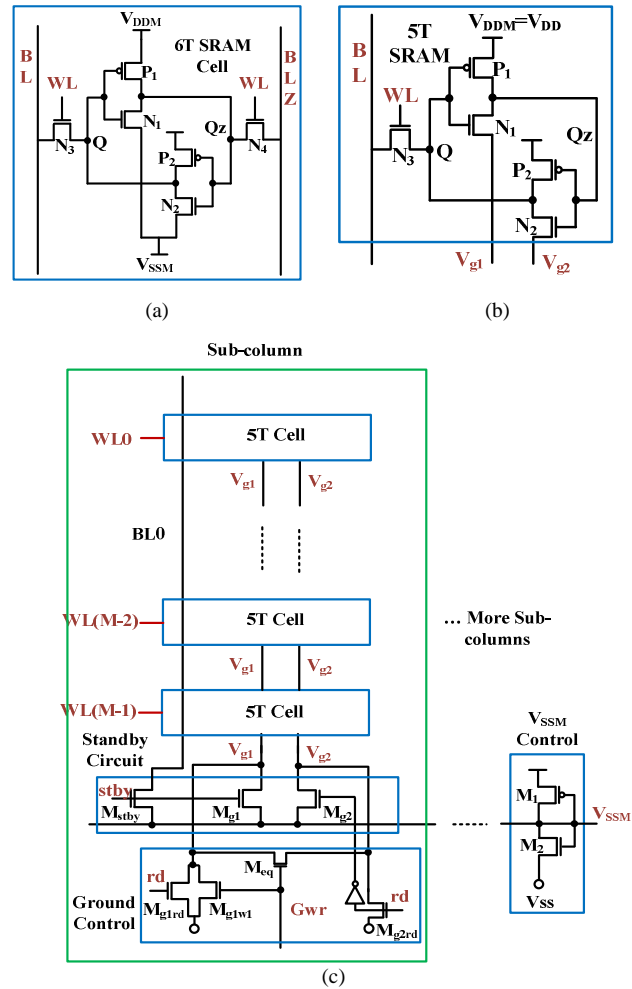

Fig. 1 (a) Conventional 6T SRAM cell, (b) Proposed 5T SRAM cell (5TSDG), (c) 5T SRAM architecture (M cells/sub-column) with sub-column circuitry and $V_{SSM}$ control.

This research is supported by CMC Microsystems and funded in part by NSERC of Canada.

TABLE I. Different types of SRAM cells used in this paper, $V_{DD} = V_{DDM}=1.3V$, $\beta = (W_{N2}/L_{N2})/(W_{N3}/L_{N3})$.

| Type | Inverter Tr.'s | Access Tr.('s) | $\beta$ | $V_{SSM}$ | BL pre-charge |
|---|---|---|---|---|---|
| *5TSDG* | HVT | SVT | 1.0 | 600mV | 600mV |
| *Low-Power 6T* | HVT | SVT | 1.4 | 600mV | $V_{DD}$ |
| *Conv. 6T* | HVT | SVT | 1.4 | 0mV | $V_{DD}$ |

## 2 5TSDG DESIGN

A conventional 6T cell in comparison with the 5TSDG is demonstrated in Fig. 1 (a) and Fig. 1 (b) respectively. A block diagram of 5TSDG cell including the sub-column circuitry is depicted in Fig. 1 (c). Standby and Ground control circuits are required one per every sub-column while $V_{SSM}$ control is shared in the entire memory array. TABLE I specifies some of the design parameters of 5TSDG, low-power 6T, as well as conventional 6T cells used in this paper for comparison. An area reduction of ~13% is predicted compared to a conventional 6T cell using standard 65nm design rules [1].

The "portless" 5T SRAM in [16] does not use a dedicated read-write port transistor, but has an "access transistor" that shorts Q and Qz nodes during read and write. $V_{DDM}$ nodes are replaced by dual bit lines for I/O and power reduction. A detailed comparison between 5TSDG and the portless 5T SRAM of [16] would be useful future work. The portless design appears to need larger PMOS and access transistors than 5TSDG.

### 2.1 Standby Mode

One of the effective and proven methods to suppress leakage power during standby in 6T SRAMs is to use dynamic sleep design while maintaining a sufficient *Static Noise Margin* (SNM), which ultimately determines the integrity of the stored data [4][6]. The most effective way to use this method is by raising the negative supply voltage of the memory cells, $V_{SSM}$, as opposed to lowering the positive one, $V_{DDM}$, to minimize bit line and cell leakage power [1][4][12].

Considering this method in 5T SRAM, a prominent feature of 5TSDG is that instead of using an external on-chip power supply to raise $V_{SSM}$ voltage above ground in standby, with existence of enough leakage sources especially sub-threshold and gate leakage currents in advanced technologies, the leaking memory array can be used as a power source to collect these charges from $V_{g1}$ and $V_{g2}$ via $M_{g1}$ and $M_{g2}$ causing a natural rise of $V_{SSM}$ to a desired biasing level using $V_{SSM}$ control circuit for fine tuning [1]. After evaluating performance, stability and power consumptions by simulations, with various combinations of threshold voltages ($V_{th}$) for each single transistor in 5TSDG, it is found that $V_{th}$ of the inverter pairs have the most significant impact on the leakage power while the access transistor, $N_3$ has the most significant impact on performance and stability. Therefore, the two inverter pairs $N_1$-$P_1$ and $N_2$-$P_2$ are selected to have high threshold voltages (HVT) while the access transistor $N_3$ has a smaller $V_{th}$, (in this case Standard $V_{th}$, SVT). All cell transistors are selected to have equal sizes ($W_i=0.15\mu m$, $L_i=0.06\mu m$).

Using two carefully sized diode-connected transistors, $M_1$ and $M_2$, the voltage across the cell in standby can be biased to remain static for various temperatures and process corners (See also [14]). In this design, a minimum voltage across the cell, $V_{min} = V_{DDM}$-$V_{SSM}$, of 0.7V is selected to yield sufficient stability [7], resulting in a simulated SNM between 181-222mV in all corners and temperatures at $V_{DDM}$=1.3V [10]. A 64Kbit memory array arranged in 64x16 blocks was simulated in standby mode using BSIM v.4 and HSPICE at $V_{DD}=V_{DDM}$=1.3V. The large capacitance of $V_{SSM}$ consisting of mostly junction and wire capacitors and sufficient available leakage current are the key factors in stability of $V_{SSM}$ during standby/write/read modes. In case of lack of leakage especially due to HVT transistors, in some corners or temperatures, $M_1$ is turned on more strongly to provide the charges to $V_{SSM}$. During read and write operations, $V_{SSM}$ remains within about 20 mV of the standby steady state value.

Another unique feature of 5TSDG that makes it different from previous research work is that $V_{SSM}$ can also be used to pre-charge the bit line, BL, in standby via $M_{stby}$ as shown in Fig. 1 (c) so that 1) channel and gate leakage through $N_3$ is reduced and minimized by up to 90% especially when a '0' is stored, and 2) the cell maintains a reasonable *Read Noise Margin* (RNM) when accessed close to the optimum achievable point and 3) To accelerate read/write operation explained in the next sections.

TABLE II shows standby leakage current and worst case RNM for various types of SRAM cells introduced in TABLE I. Fig. 2 compares the power consumption of 5TSDG including peripheral circuits with a low-power 6T design in various process corners. Traditional 5T designs as in [8][9], where $V_{SSM}$ is held at $V_{SS}$ level, require lower $V_{th}$ for internal cell transistors in 65nm technology, such as $N_1$, and $P_2$ (Fig. 1), to enable write '1' operation discussed in section 2.4. Thus, even though some leakage power is saved by cutting a bit line and biasing the other to a lower voltage, the overall leakage is quite high, being about half of the conventional 6T cell value in TABLE II.

TABLE II. Leakage current and RNM comparison in different SRAM types (not including peripheral circuits)

| 64 cells | *5TSDG* | *Low-Power 6T* | *Conventional 6T* |
|---|---|---|---|
| *Leakage (nA)* | 80.6 | 88.7 | 2020.0 |
| *RNM (mV)* | 172.3 | 123.2 | 123.2 |

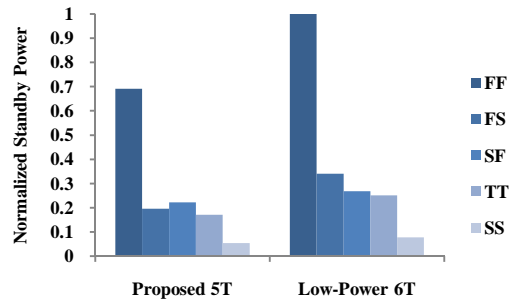

Fig. 2 5TSDG vs. low-power 6T SRAM designs: (64Kb SRAM array and peripheral circuits, FF corner on the left, 120°C, 50% '0's 50% '1's stored in the array, 1≈1.3mW)

## 2.2 Read Operation

The read operation is similar to a 6T SRAM except that only one bit line is used. In 5TSDG, the bit line is pre-charged in standby by $V_{SSM}$ which is near the optimum point to maximize RNM in the worse case (FS). Another advantage of this pre-charge method compared to [9] is that it does not require an additional power supply on chip such as a DC-DC converter or a level shifter which will add to the chip area and power consumption. A simple sense amplifier circuit used in 5TSDG is shown in Fig. 3 (b). Although not the fastest type, it is attractive due to its simplicity and that it does not need a clock signal [13]. During read, rd signal in Fig. 1 (c) is raised causing $V_{g1}$ and $V_{g2}$ to be pulled down to $V_{SS}$ by $M_{g1rd}$ and $M_{g2rd}$, which will maximize RNM and read performance. The global bit line, Gbit, is the output of the sense amplifier and is pre-charged to $V_{SSM}$ through $M_8$ in standby and is pulled down to $V_{SS}$ by $M_7$ during a read '0'. Therefore, a read '1' is always implied unless Gbit is pulled down. Inverter $M_5$-$M_6$ should have a sufficient noise margin to prevent a false trigger. This sense amplifier can be shared by two bit lines from two adjacent sub-columns. For instance, in a 128-cell column composed of two 64 cell sub-columns, the sense amplifier is placed in between bit lines BitL and BitR. SelL and SelR signals should be selected by a row decoder to select the appropriate bit line to read from. $M_3$ and $M_4$ are used to pre-charge the input of the inverter $M_5$-$M_6$ in standby. For similar bit line capacitances, read speed in 5T and 6T SRAMs is comparable.

Fig. 4 (a) shows simulation results of the read operation in a conventional 6T cell using a sense amplifier shown in Fig. 3 (a). In this simulation, WL pulse is artificially generated such that BL reaches about $V_{DD}/2$ in read for power saving reasons. Gbit and Gbitz are the outputs of the sense amplifier, and are pre-charged high using prez pulses before the read operation.

Fig. 4 (b) demonstrates the read operation of 5TSDG using a sense amplifier shown in Fig. 3 (b) in a 64Kbit memory array arranged in 64x16 bit blocks for two neighboring cells sharing the same word line, WL, storing a '0' and a '1' on Q0 and Q1 nodes, and having two bit lines BL0 and BL1 respectively. Gbit load in 5TSDG is the same as that in Gbit and Gbitz in the 6T counterpart. Q0-Q0z-BL0-Gbit0 and Q1-Q1z-BL1-Gbit1 are related to a cell in 5TSDG storing a '0' and '1' respectively and sharing the same word-line (WL).

A dynamic increase in Q0 node occurs while reading a '0' due to the current flow from the bit line to $N_3$ and $N_2$ as the word line is raised and as shown in Fig. 4 (b) ($Q_{max}$). During read '1' a drop of voltage in Q1 node is observed for a similar reason ($Q_{min}$). $Q_{max}$ and $Q_{min}$ should not cause a read upset i.e. they should be less and more than the tripping voltage of the inverter pairs, respectively, to avoid turning a read into a write, especially in FS corner. To further reduce the probability of read-upset in 5T cell, it is possible to increase word-line rise time and make the bit lines shorter to reduce their capacitances [2]. The latter may also improve read speed by reducing bit line swing delay.

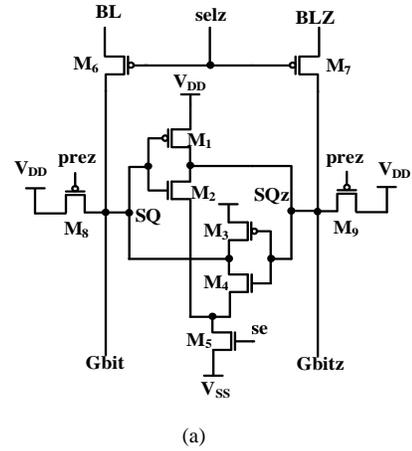

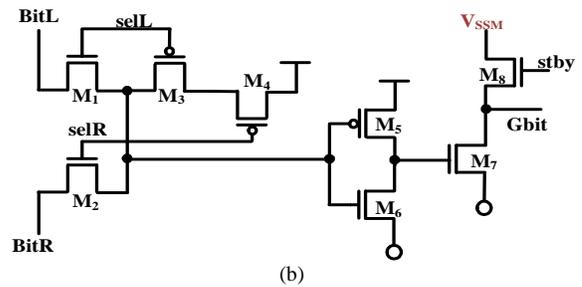

(a)

(b)

Fig. 3 Sense amplifier used in the read operation of the conventional 6T (a) and 5TSDG cells (b).

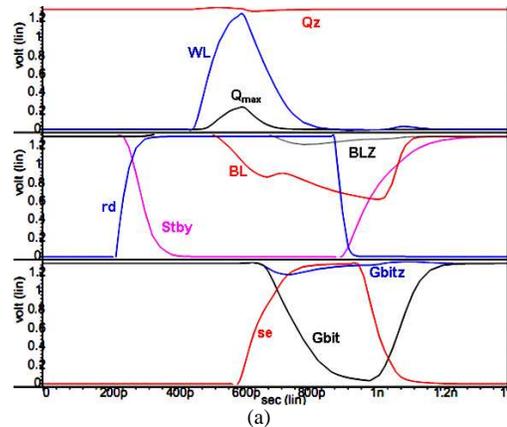

(a)

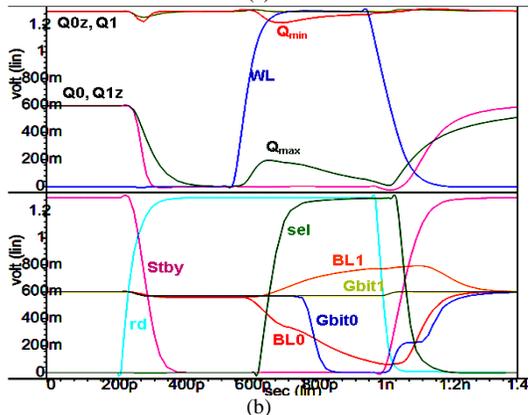

(b)

Fig. 4 (a) Read operation in a conventional 6T SRAM (b) Read operation in 5TSDG, (typical corner (TT), 120°C).

TABLE III shows the trip voltage vs. Qmax and Qmin while reading in different corners in 5TSDG. These results when compared with RNM measurements for various mismatch cases in worst case corner FS, show the stability of 5TSDG. RNM is measured with $V_{SSM}$ on the bit line similar to [9]. The best biasing value for $V_{SSM}$ maximizes RNM for the FS corner.

TABLE III. 5TSDG: Trip point voltage vs. $Q_{max}$ and $Q_{min}$, 120°C)

| 5T Cell Trip Voltage (mV) /$Q_{max}$ (mV) | | | | |
|---|---|---|---|---|
| FF | **FS** | SF | TT | SS |
| 883/193 | **866/185** | 934/202 | 899/195 | 905/196 |
| 5T Cell $Q_{min}$ (V) | | | | |
| 1.19 | **1.18** | 1.24 | 1.22 | 1.24 |

## 2.3 $V_{SSM}$ Stability in Dynamic Mode

During standby mode of the 5TSDG cell, $V_{SSM}$ is used as a power supply to raise $V_{g1}$ and $V_{g2}$ above $V_{SS}$ and pre-charge the bit-line. During read operation, $V_{g1}$ and $V_{g2}$ are driven to $V_{SS}$ to maximize RNM and the read speed. On the other hand, after a read operation is completed, $V_{g1}$, $V_{g2}$ and the bit-line are driven back to $V_{SSM}$ since the memory cell will be in standby again. This voltage swing of $V_{g1}$, $V_{g2}$ and the bit-line affects voltage level of $V_{SSM}$ since each re-charge of these voltages takes charges away from $V_{SSM}$ causing it to drop by an amount of $\Delta V_i$, where *i* is the index of consecutive read operations. In a case of reading a '1', the bit line will actually add charges to $V_{SSM}$ but that amount is much less than the effect of the ground lines taking away charges after being driven low for a read. Fortunately, $V_{SSM}$ is highly capacitive with much higher capacitance than $V_{g1}$ and $V_{g2}$, and many memory cells in standby provide electric charges to it. Therefore, $V_{SSM}$ changes very little during read operation especially when it has large capacitance (attached to large memory arrays), and even if it does, it will actually help the read operation in terms of performance and read noise margin (see Fig. 5). In addition, $V_{SSM}$ does not decrease beyond a steady-state value, and when reading is complete, it is pulled back towards its standby level due to an increase in memory cell leakage (see Fig. 6 and Fig. 7). Fig. 8 shows how $V_{SSM}$ reaches a steady-state value after many read operations for different SRAM array sizes (64Kb, 1Mb, and 2Mb) in FF corner. This figure demonstrates that when larger number of memory cells are attached to $V_{SSM}$, the initial values of $\Delta V_i$ which are instantaneous voltage decays after each read, and the total decay to reach the steady-state value, $\Delta V_{tot}$, will be smaller than that of smaller arrays. After each read cycle, $\Delta V_i$ is reduced until it reaches 0V. At this point (steady state), the memory leakage is sufficiently increased such that it can fully replenish the lost charges between read cycles. $V_{SSM}$ voltage after each read cycle (*i*) can be described by equation 1.

$$V_{SSM}(i+1) \cong \varphi(i).V_{SSM}(i) + \frac{i_{m_{avg}}(i).\Delta t}{C_{VSSM}(stby)} \quad (1)$$

where $\varphi(i) = \frac{C_{V_{SSM}}(read) + (\frac{V_{BL_0}}{V_{SSM}(i)}N_{B_0} + \frac{V_{BL_1}}{V_{SSM}(i)}N_{B_1})C_{BL}}{C_{VSSM}(stby)}$,

$C_{VSSM}(stby) = C_{V_{SSM}}(read) + C(SubCol)$,

$C(SubCol) = (N_{B_0} + N_{B_1})(C_{BL} + C_{V_{g1}} + C_{V_{g2}})$,

$C_{V_{g1}}$, $C_{V_{g2}}$, $C_{BL}$, and are the capacitances of $V_{g1}$, $V_{g2}$, and the bit-line respectively. It is assumed that $V_{g1}$ and $V_{g2}$ have been driven to 0V initially. $V_{BL_0}$ and $V_{BL_1}$ are the bit-line voltages after a read '0' and a read '1' respectively. $N_{B_0}$ and $N_{B_1}$ are the number of '0' and '1' bits in a word respectively (16 bits/word in the simulation results of this paper). In standby, $C_{VSSM}(stby)$ includes the capacitance of all sub-column connections, and $V_{SSM}$ interconnections. During read, a single sub-column with capacitance of $C(SubCol)$ is removed. $i_{m_{avg}}(i)$ is the average memory leakage current over the i-th read cycle period, $\Delta t$. It is increased as $V_{SSM}$ is reduced. Similarly, $V_{SSM}(i)$ is the $V_{SSM}$ voltage at the end of the i-th read cycle. As the memory array size is increased, $\varphi(i)$ approaches to one since $C_{V_{SSM}}(read)$ approaches $C_{VSSM}(stby)$. Part of $\Delta V_i$ is caused by a small amount of overlap between rd and stby signals in Fig. 4 (b).

This effect on $V_{SSM}$ occurs also in write operation when the bit-lines are charged and discharged. However, for explanatory purposes, read operation, which is the most severe, is selected to be demonstrated. The number of cells per bit-line and number of bits per word also contribute to this effect.

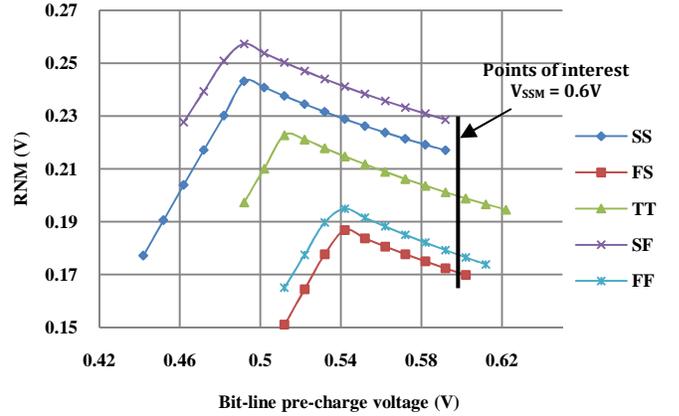

Fig. 5 RNM variations vs. bit-line biasing (=$V_{SSM}$, e.g. vertical line) for various corners in the 5TSDG cell

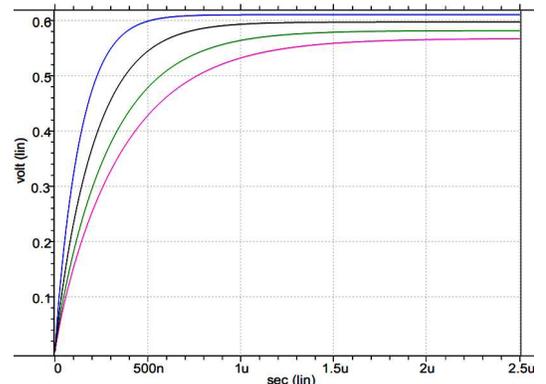

Fig. 6 The response of VSSM when forced to '0' volts and left floating in standby for 64Kb, 128Kb, 256Kb, and 512Kb from left to right for 5T SRAM array (FF corner, 120°C)

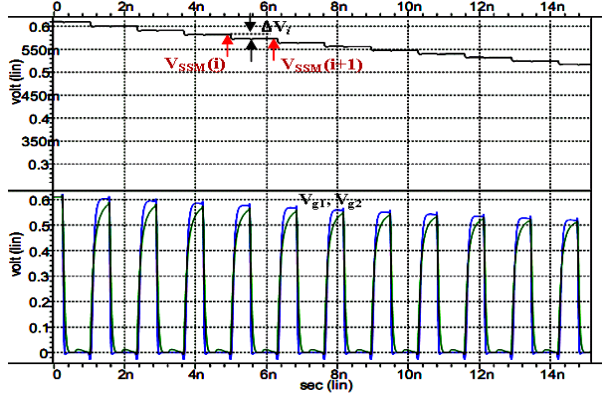

Fig. 7 Effect of read operation on $V_{SSM}$ (64Kb array 16 bits/word, FF, 120°C, $\Delta t \approx 1.4$ns)

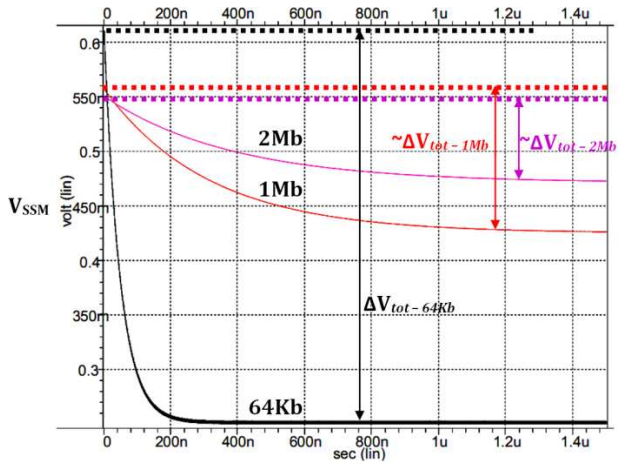

Fig. 8 $V_{SSM}$ Saturation during repeated read operations for 64Kb, 1Mb and 2Mb 5T SRAM arrays (64 bits/word, FF, 120°C)

### 2.4 Write Operation

Since a 5T SRAM cell only has a single bit-line, writing either a '0' (W0) or a '1' (W1) into the cell is performed using the same bit-line. This is different from the 6T structure where there is technically no difference between a W0 or a W1, i.e. by selectively pulling down one of the bit lines depending on the data status, a W0 operation is applied on one side of the cell and the feedback will recover the opposite storage node to the complement value. In 5TSDG, W0 is performed in a similar way. On the other hand, in W1, the bit-line is pulled high by global write signal, Gwr, so that when the word-line is selected, state toggle is initiated. Gwr is driven high by the write circuit in W1 and is driven to $V_{SS}$ otherwise. Using conventional 6T transistor ratios and sizing, it is almost impossible to write a '1' in a 5T cell because in a 6T cell: 1) $N_2$ needs to be stronger than $N_3$ by *cell ratio* (CR) factor $\beta$, typically between (1.2~1.5) to maintain read stability [4]. 2) $P_1$ and $P_2$ need to be weak enough, usually minimum size for *write-ability* purposes. 3) The access transistor is an NMOS, which does not pull up strongly due to its physical nature. These constraints will oppose raising Q if applied in a 5T memory cell for a W1 using a single bit line. To combat this problem, [9] suggests using different (W/L) sizes for the transistors such as, using a CR of ~0.45, weakening $P_1$, strengthening $P_2$ and $N_1$ with the cost of noise margin. As opposed to 5TSDG in this paper, design in [9] will cause a 50% reduction of RNM when compared to conventional 6T cell and therefore is more susceptible to performance fluctuations in more advanced technologies, especially due to process variations.

On the other hand, to make W1 possible, [8] suggests disconnecting $Vg_2$ from $V_{SS}$ and letting it float near a biasing voltage by using a capacitor while keeping $Vg_1$ at $V_{SS}$ during write. This method will weaken $N_2$ by lowering its $V_{DS}$ which will facilitate W1. However, this method does not take advantage of leakage power reduction opportunities.

As illustrated in Fig. 1 and as discussed earlier, in 5TSDG, $V_{SSM}$ is connected to $Vg_1$, $Vg_2$ and the bit lines in standby mode. In W0, $Vg_2$ stays connected to $V_{SSM}$ via $Mg_2$ while $Vg_1$ floats near $V_{SSM}$. In W1, $Vg_1$ is pulled down to $V_{SS}$ through $M_{g1w1}$. $M_{equ}$ is turned on by Gwr signal which is high when W1 and is at $V_{SS}$ otherwise. The role of this transistor is to limit $\Delta Vg=Vg_2-Vg_1$ as shown in Fig. 9 to improve SNM of the disturbed cells in the same sub-column. The strength of $M_{equ}$ is chosen through simulation to limit write disturb for all process corners especially for fast NMOS corner cases [2]. This disturbance can also be minimized by reducing the write pulse period to its limit. In summary, in W1, $N_1$ will have a stronger current drive than $N_2$ since its $V_{DS}$ is maximized i.e. increased by $V_{SSM}$.

The threshold voltage of access transistor, $N_3$, plays a key role in W1 performance. Simulation results reveal that standby power varies less than 2% using high, standard or low $V_{th}$ (HVT, SVT, LVT) for $N_3$. In order to improve W1 performance, the $V_{th}$ of $N_3$ can be reduced with some loss of RNM. In 5TSDG, $V_{th}$ of $N_3$ can be between the HVT and LVT to maintain a reasonable RNM/W1 performance as shown in TABLE IV (for W1 delay measurement see Fig. 9). RNM can be further increased by reducing bit-line capacitance and/or increasing word-line rise time [2].

TABLE IV. RNM and W1 comparison for different $V_{th}$ for $N_3$, at worst case RNM (FS corner, 120°C)

| 5T Cell RNM (mV) / W1 Delay (ps) for Various N3 $V_{th}$ | | |
|---|---|---|
| LVT (~230 mV) | SVT (~440 mV) | HVT (~600 mV) |
| 144.1/96.8 | 172.3/116.4 | 225.1/170.6 |

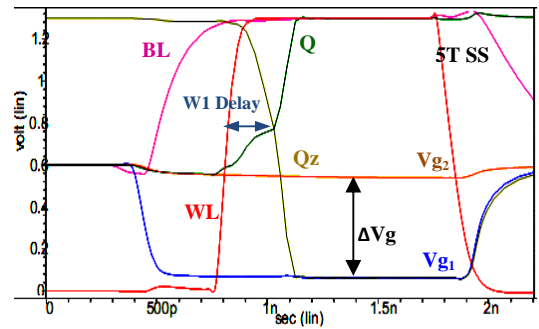

Fig. 9 W1 operation of 5TSDG cell in slow corner (SS) (120°C)

Therefore, to improve read stability and write-ability (particularly W1), the solution is to find a reasonable mid-point considering the fact that $N_3$ does not play a key role in standby power consumption. Limited to three choices for $V_{th}$ selection, SVT for $N_3$ is reasonable as shown in TABLE IV. However, in chip foundries, even a lower threshold somewhere between LVT and SVT can be achieved by changing gate oxide thickness. Fig. 11 compares W0 and W1 performance of 5TSDG with a low-power 6T SRAM described in TABLE I. For both cases, W1 delay is measured from when WL = 50%$V_{DDM}$ to when Q=80%$V_{DDM}$, and W0 delay is measured similarly but when Q=20%$V_{DDM}$ above $V_{SS}$. This measurement is different from what was reported in TABLE IV (word-line to Q-Qz cross point). W1 can be ~11-31% slower than a conventional 6T design and can be improved by reducing $V_{th}$ of $N_3$. W0 performance is similar to conventional 6T cell.

Fig. 10 (a) shows how the voltage of $V_{g1}$ in W1 can affect SNM on disturbed cells while driving $V_{g2}$ at a fixed voltage (at $V_{SSM}$) mimicking that there is no $M_{equ}$. Fig. 10 (b) demonstrates the reverse scenario where $V_{g1}$ is fixed at 0V, and $V_{g2}$ varies from 0V to $V_{SSM}$. Similarly, this figure shows that with no weak equalization between $V_{g1}$ and $V_{g2}$, the disturbed cells are susceptible to data corruption due to environmental disturbances. The strength of $M_{equ}$ will determine the limitation on this disturbance by both lowering $V_{g2}$ from $V_{SSM}$ and not allowing $V_{g1}$ to be pulled down so much. In 5TSDG, $M_{equ}$ was ratioed such that the W1 disturbed SNM was greater than ~50mV.

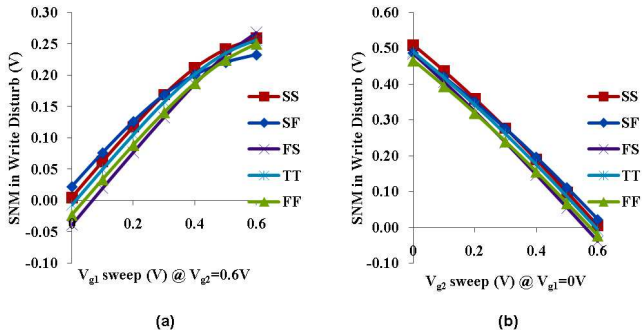

Fig. 10 SNM in W1 disturb cells vs. $V_{g1}$ and $V_{g2}$ voltages (120℃)

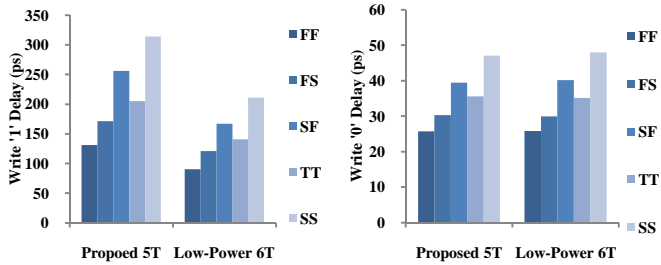

Fig. 11 W1 (left) and W0 (right) delay comparison in different corners between 5TSDG and low-power 6T cell (120℃)

The write margin of the proposed 5T SRAM design can be divided into W0 margin (W0M), and W1 margin (W1M) since as opposed to the 6T cell counterpart, W0 and W1 have different WMs. One of the common methods to measure WM in conventional 6T SRAMs is by measuring the maximum BL voltage able to flip the cell state [15]. For 5TSDG, W1M is defined to be the difference between the positive supply voltage, $V_{DDM}$, and the minimum BL voltage able to write a '1' into the cell while W0M is defined to be the maximum BL voltage able to write a '0' into the cell. In the 5TSDG ($V_{DDM}$=1.3V), for a typical-typical corner (TT), W1M is ~0.5V, and W0M is ~0.4V.

## 3 DYNAMIC POWER CONSUMPTION

Dynamic power consumption of 5TSDG can be divided into read and write power. Power consumption during read is a function of Vmin, which determines $V_{SSM}$ biasing level. During several consecutive reads, $V_{g1}$ and $V_{g2}$ in Fig. 1 are driven to $V_{SS}$ and $V_{SSM}$ frequently. Active power consumption is changed as supply voltage is changed due to the square law dependency. This power is also dependent on the frequency of $V_{SSM}$ swing during read. Equation 2 shows the dynamic power consumed due to the voltage swing of the ground lines of 5TSDG, where $C_L$, is the summation of $V_{g1}$ and $V_{g2}$ capacitances, $\Delta V$ is $V_{SSM}$-$V_{SS}$ and $f$ is the frequency of voltage swing.

$$P_{dyn} = C_L \Delta V^2 f \qquad (2)$$

Reading a '0' (R0) consumes more power than reading a '1' (R1) since in R0, the bit-line is pulled sufficiently low to trigger the sense amplifier, and the global bit-line of the sense amplifier is also pulled down. In R1, bit-line is only required to be pulled high enough to avoid activating the sense amplifier, and the global bit-line stays at $V_{SSM}$. Fig. 12 shows read power and standby power for various $V_{DDM}$ values while keeping Vmin=$V_{DDM}$-$V_{SSM}$ constant at 0.7V for a 64x16 bit block of 5TSDG. As $V_{DDM}$ is increased, $V_{SSM}$ also increases accordingly causing $\Delta V$ in equation 2 to increase during read operation. Therefore, read power is increased quadratically with higher $V_{DDM}$.

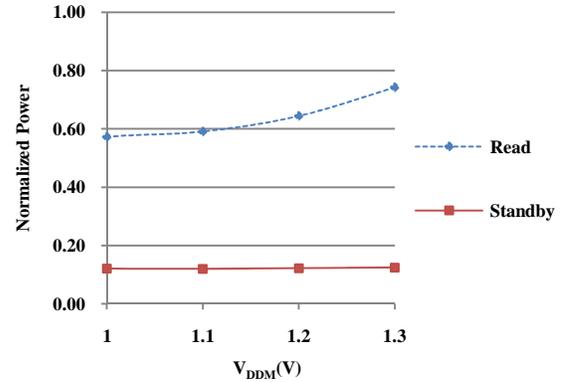

Fig. 12 Comparison of normalized read and standby power vs. $V_{DDM}$ for 5TSDG cell, 64x16 bit block, reading 16 '0's continuously from 16-bit words (FF corner, 120℃)

Fig. 13 demonstrates case study results of worst-case (FF, 120°C) normalized power consumption in standby mode, read, and write operations of 5TSDG in comparison with low-power 6T design. Other corners have similar results comparable to Fig. 2. Read power consists of standby power of the idle memory cells, and the dynamic power described by equation 2. In this case study where a 64Kbit array consisting of 64x16 bit blocks was studied (reading continuously from a 16-bit word), 5TSDG could achieve up to ~30% power reduction in read mode compared to that of the low-power 6T structure. In this example, R1 consumes ~7% less power in 5TSDG compared to a R0 as explained earlier. Obviously, larger number of read operations will result in a linearly higher power consumption difference in comparison with standby power due to larger values of $f$ in equation 2. Read operation of the low-power 6T and 5TSDG designs in this experiment were similar to Fig. 4(a) and Fig. 4(b) respectively. In a pipelined "smart" memory, back-to-back reads from the same sub-block would consume less dynamic power if $V_{g1}$ and $V_{g2}$ are held at $V_{SS}$ between consecutive reads.

The 5TSDG write power can be divided into W0 and W1 power, each consisting of idle cell standby power, plus the dynamic power. In Fig. 13, a 64Kbit array consisting of 64x16 bit blocks was studied while writing into a 16-bit word. In this example, W0 consumes ~80% less power, and W1 consumes ~9% less power compared with a low-power 6T structure in worst case scenario (FF corner, 120°C). Since R0 and R1 use similar power, storing bits to favor W0 (i.e. cell inverted) may reduce total power.

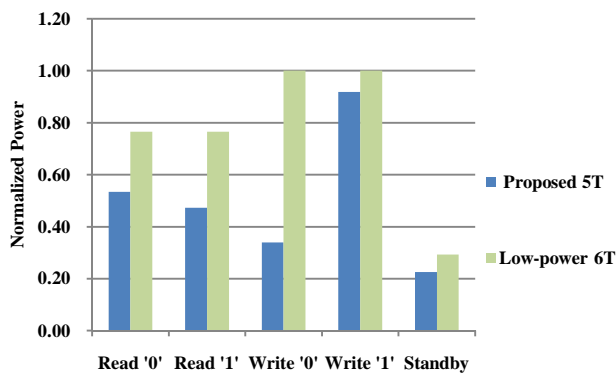

Fig. 13 Case study results of the worst-case write power consumption in comparison with read and standby power for 5TSDG vs. low power 6T design (FF, 120°C), 1≈ 33.8mW

## 4 CONCLUSION

In this paper, the operation of a new low-power and high performance design for a 5T SRAM cell was addressed which has improvements in static and dynamic power consumption, stability margins and performance when compared to previous designs in this area. The stability of the novel biasing scheme in dynamic mode was analyzed. The reduction in dynamic power consumption in comparison with a low-power 6T counterpart was demonstrated. A significant area saving is predicted compared to a conventional 6T cell.